\documentclass[aps,prl,twocolumn,superscriptaddress,showpacs]{revtex4}

\usepackage{graphics}
\usepackage{amsmath}
\usepackage{bm}

\begin{document}

\title{Spectral functions of strongly interacting isospin-$\frac12$ bosons in
  one dimension}

\author{K. A. Matveev}

\affiliation{Materials Science Division, Argonne National Laboratory, 
Argonne, Illinois 60439, USA}

\author{A. Furusaki}

\affiliation{Condensed Matter Theory Laboratory, RIKEN, Wako, Saitama
  351-0198, Japan}

\date{August 12, 2008}

\begin{abstract}
  We study a system of one-dimensional (iso)spin-$\frac12$ bosons in the
  regime of strong repulsive interactions.  We argue that the low-energy
  spectrum of the system consists of acoustic density waves and the spin
  excitations described by an effective ferromagnetic spin chain with a
  small exchange constant $J$.  We use this description to compute the
  dynamic spin structure factor and the spectral functions of the system.
\end{abstract}


\maketitle

Physics of one-dimensional Fermi systems has long attracted the interest
of both theorists and experimentalists.  Interactions between particles
have a strong effect on the properties of these systems.  Interacting
fermions form the so-called Luttinger-liquid state \cite{giamarchi}, whose
excitations are bosons with acoustic spectrum, $\varepsilon(q)\propto
|q|$.  Recently it has become possible to confine ultracold gases of
bosons to elongated traps \cite{kinoshita, paredes}, effectively creating
systems of one-dimensional bosons.  The properties of interacting
one-dimensional spinless bosons are in many respects similar to those of
spinless fermions.  In particular, they too form a Luttinger-liquid state
at low energies.

In a recent experiment \cite{mcguirk} bosons with two internal degrees of
freedom, which can be viewed as components of (iso)spin-$\frac12$, were
confined to one dimension.  For spin-$\frac12$ particles the difference
between the Bose and Fermi statistics is of fundamental importance.
Indeed, spin-independent interactions between one-dimensional bosons favor
ferromagnetic spin ordering \cite{eisenberg}, whereas for fermions the
ground state spin is zero \cite{lieb}.  As a result, the low-energy spin
excitations of the boson system are magnons with quadratic spectrum
$\varepsilon(q)\propto q^2$, and the system is no longer a Luttinger
liquid.

In the absence of the effective theory of interacting spin-$\frac12$
bosons in one dimension, considerable progress has been made recently by
focusing on the regime of very strong repulsive interactions
\cite{akhanjee, zvonarev}.  In this paper we show that this regime allows
for a remarkably simple theoretical description, in which there are two
types of low energy excitations: acoustic density waves and the spin
excitations described by a one-dimensional Heisenberg model with a very
small ferromagnetic exchange constant $J$.  The theory is applied to the
calculation of the dynamic spin structure factor and the spectral
functions of the system.  Unlike Refs.~\onlinecite{akhanjee} and
\onlinecite{zvonarev}, our conclusions are not limited to spin excitations
of small momentum $q\to0$.  In addition, although the frequency $\omega$
is assumed to be small compared to the typical kinetic energy of the
bosons $E_F\sim (\hbar n)^2/m$, it can be of order of the small exchange
constant $J$.  (Here $n$ is the one-dimensional density of bosons and $m$
is their mass.)

The model we consider is that of one-dimensional (iso)spin-$\frac12$
bosons interacting with repulsive spin-independent potential $V(x-y)$.
For simplicity, we concentrate on the most realistic regime of short-range
interactions, $V(x-y)=g\delta(x-y)$; the generalization to the case of
finite-range repulsion is relatively straightforward.  The strong
repulsion regime is achieved at $\gamma\gg1$, where $\gamma=mg/\hbar^2 n$
is the dimensionless interaction strength.  

As the first step, we show that at low energies the excitation spectrum of
the system consists of independent phonon and magnon excitations.  This
effect is essentially equivalent to the well-known spin-charge separation
in interacting one-dimensional electron systems \cite{dzyaloshinskii}.
Our arguments follow the discussion \cite{matveev, MFG} of that phenomenon
in the limit of strong repulsion.

In the Tonks-Girardeau limit $\gamma\to+\infty$ the repulsion effectively
forbids any two particles to occupy the same point in space, regardless of
their spin.  Thus the density excitations of the system are those of a gas
of spinless hard-core bosons, or, equivalently, those of non-interacting
gas of spinless fermions \cite{tonks-girardeau}, where the same constraint
is enforced by the Pauli principle.  It is convenient to treat the
low-energy excitations of one-dimensional spinless Bose and Fermi systems
in the framework of the hydrodynamic approach \cite{giamarchi} and write
the Hamiltonian in the form
\begin{equation}
  \label{eq:H_ph}
  H_{ph} = \frac{\hbar u_\rho}{2\pi}\int\left[K (\partial_x\theta)^2 
                         +K^{-1}(\partial_x\phi)^2\right]dx.
\end{equation}
Here $\phi$ and $\theta$ are bosonic fields satisfying the standard
commutation relation $[\phi(x),\partial_y\theta(y)]=i\pi \delta(x-y)$.
The Luttinger liquid parameter $K$ and the phonon velocity $u_\rho$ are
determined by the interactions.  In the case of hard-core bosons $K=1$,
while the effective ``Fermi velocity''  $u_\rho=\pi\hbar n/m$.

In the limit $\gamma\to+\infty$ any collision of two bosons results in
perfect backscattering.  As a result the bosons become distinguishable
particles.  Indeed, if boson 1 is to the left of boson 2, i.e., $x_1<x_2$
at some moment in time, then this property cannot be changed as a result
of any collisions between particles.  Thus one can number all particles by
an integer $l$ in accordance with their positions along the $x$-axis.  In
this limit the spins of the bosons do not interact, and each state of $N$
bosons is $2^N$-fold degenerate.  A coupling of the spins appears only
when $\gamma$ is finite.  At $\gamma\gg1$ a collision of two bosons, $l$
and $l+1$, may result in their forward scattering, in which case the
particles exchange their spins.  Since for spin-$\frac12$ particles the
spin permutation operator $P_{l,l+1}$ can be expressed as $P_{l,
  l+1}=2{\bm S}_{l}\cdot{\bm S}_{l+1}+1/2$, this gives rise to coupling of
the spins of the nearest-neighbor particles:
\begin{equation}
  \label{eq:Heisenberg}
  H_\sigma = -\sum_l J\,{\bm S}_{l}\cdot{\bm S}_{l+1}.
\end{equation}
Thus at $\gamma\gg1$ the low-energy excitations of the system are given by
the acoustic phonons, described by the Hamiltonian (\ref{eq:H_ph}) and the
spin excitations of the Heisenberg Hamiltonian (\ref{eq:Heisenberg}).  

A similar separation of the density and spin excitations is well known in
the case of strongly interacting one-dimensional fermions, where it was
first derived \cite{ogata} from the exact solution of the infinite-$U$
Hubbard model.  The sign of the exchange constant $J$ is determined by the
requirement to either symmetrize or antisymmetrize the wave function with
respect to the permutation $x_l\leftrightarrow x_{l+1}$; the coupling is
antiferromagnetic for fermions, $J<0$, and ferromagnetic for bosons,
$J>0$.  On the other hand, the magnitude of the exchange constant $J$ is
determined by the amplitude of the forward scattering of two neighboring
particles, regardless of their statistics.  Thus we find the same value of
$J$ as in the case of fermions with strong short-range repulsion,
\begin{equation}
  \label{eq:J_point-like}
  J = \frac{2\pi^2}{3}\frac{\hbar^2n^2}{m\gamma},
\end{equation}
see Eq.~(22) of Ref.~\onlinecite{matveev}.  The effective theory
(\ref{eq:Heisenberg}), (\ref{eq:J_point-like}) of the spin subsystem is
consistent with the recent thermodynamic Bethe ansatz results \cite{guan}.

The Hamiltonian describing all the low-energy excitations of the system is
the sum $H_{ph}+H_\sigma$.  An important assumption in its derivation was
that all the relevant energy scales in the problem, such as the
temperature $T$, are small compared to the bandwidth (the ``Debye
frequency'') of the phonons $E_F$.  In the following we limit our
discussion to the most interesting case of $T=0$.

The ground state of the ferromagnetic spin chain (\ref{eq:Heisenberg}) is
fully spin-polarized.  The excitations near this state, the magnons, have
the well-known spectrum
\begin{equation}
  \label{eq:magnon_spectrum}
  \varepsilon(Q) = J(1-\cos Q),
\end{equation}
where $Q$ is the wave vector defined with respect to the lattice of the
spin chain (\ref{eq:Heisenberg}) and varying in the range $-\pi<Q<\pi$.
Since the spins are attached to particles filling the real space with
density $n$, the physical momentum of the magnon is $p=\hbar nQ$
\cite{brazovskii, penc1}.  In the limit of small $p$, the spectrum
(\ref{eq:magnon_spectrum}) is quadratic, $\varepsilon(p)=p^2/2m^*$.  Using
Eq.~(\ref{eq:J_point-like}), one finds the effective mass
$m^*=(3/2\pi^2)\gamma m$, in agreement with the result of
Ref.~\onlinecite{fuchs}.

Let us now illustrate our approach based on the separation of the density
and spin excitations in the form (\ref{eq:H_ph}) and (\ref{eq:Heisenberg})
by calculating the dynamic spin structure factor
\begin{equation}
  \label{eq:structure_definition}
  S_\perp(q,\omega) = \int\frac{dx\,dt}{2\pi} e^{-iqx+i\omega t}
                     \langle
                         S^+(x,t) S^-(0,0)
                     \rangle.
\end{equation}
Here $\bm{S}(x)$ is the spin density operator, $S^{\pm}=S^x\pm i S^y$, and
the expectation value $\langle\ldots\rangle$ is evaluated in the fully
polarized ground state of the system, with the polarization assumed to be
directed in the positive $z$-direction.

We start by expressing the spin density operator $\bm{S}(x)$ in terms of
the particle density operator $n(x)$ and the spin operator $\bm S_l$,
\begin{equation}
  \label{eq:spin_density_operator}
  \bm S(x) = n(x) \bm S_{l(x)}.
\end{equation}
Here $l(x)$ is the operator of the number of particles to the left of
point $x$, i.e., $\partial_x l(x)=n(x)$.  Its presence in
Eq.~(\ref{eq:spin_density_operator}) accounts for the fact that the
operator $\bm{S}(x)$ acts on site $l$ of the spin chain
(\ref{eq:Heisenberg}) attached to the boson at point $x$,
cf.~\cite{MFG}.

The problem of zero-temperature properties of strongly interacting bosons
is considerably simpler than that of fermions \cite{MFG}, because of the
simplicity of the ground state of the ferromagnetic Heisenberg model
(\ref{eq:Heisenberg}) and its single particle excitation spectrum
(\ref{eq:magnon_spectrum}).  In particular, the correlator $\langle S_l^+
S_{l'}^-\rangle_\sigma$ for the spin chain (\ref{eq:Heisenberg}) is easily
found as
\begin{equation}
  \label{eq:spin-spin}
  \langle S_l^+(t) S_{l'}^-(0)\rangle_\sigma 
        = \int\frac{dQ}{2\pi}\, e^{iQ(l-l')-i\Omega(Q)t},
\end{equation}
where $\Omega(Q)=\varepsilon(Q)/\hbar$ is given by
Eq.~(\ref{eq:magnon_spectrum}).  Substituting
Eq.~(\ref{eq:spin_density_operator}) into (\ref{eq:structure_definition})
and using (\ref{eq:spin-spin}), we find
\begin{eqnarray}
  \label{eq:structure_intermediate}
  S_\perp(q,\omega) &=& \int\frac{dx\,dt\,dQ}{(2\pi)^2}\,
                      e^{-iqx+i[\omega-\Omega(Q)] t}
\nonumber\\
                 &&\times\left\langle 
                       e^{iQ[l(x,t)-l(0,0)]}n(x,t)n(0,0)
                         \right\rangle_{ph}.
\end{eqnarray}
The expectation value $\langle\ldots\rangle_{ph}$ is performed in the
ground state of the phonon Hamiltonian (\ref{eq:H_ph}).  To evaluate it,
we use the standard hydrodynamic expression for particle density $n(x) = n
+\frac{1}{\pi}\partial_x\phi(x)$, and the resulting expression for the
particle number
\begin{equation}
  \label{eq:l_bosonized}
  l(x) = nx +\frac{1}{\pi} \phi(x).
\end{equation}
In the low-energy limit one can neglect the $\partial_x\phi$ correction to
$n(x)$, and replace it with the average value $n$.  However, it is
important to include the field $\phi$ in Eq.~(\ref{eq:l_bosonized}) when
evaluating the exponential in the second line of
Eq.~(\ref{eq:structure_intermediate}).  The latter calculation is
performed using the standard techniques \cite{giamarchi}, resulting in
\begin{equation}
  \label{eq:standard_bosonization}
  \left\langle 
      e^{iQ[l(x,t)-l(0,0)]}
  \right\rangle_{ph}
  =\frac{e^{inQx}}{\left[(1+iDt)^2 +
    (Dx/u_\rho)^{2}\right]^{(Q/2\pi)^2}},
\end{equation}
where $D\sim E_F/\hbar$ is the phonon bandwidth.  

In the denominator of Eq.~(\ref{eq:standard_bosonization}) one can neglect
$x$ compared to $u_\rho t$.  Indeed, to this approximation one finds that
Eq.~(\ref{eq:standard_bosonization}) falls off at $Q\sim
1/\sqrt{\ln(Dt)}$, resulting in the estimate $x\sim 1/nQ\sim
\sqrt{\ln(Dt)}/n\ll u_\rho t$, cf.~\cite{zvonarev,fiete}.  The remaining
calculation is straightforward, and one finds
\begin{equation}
  \label{eq:structure_result}
  S_\perp(q,\omega) =
      \frac{\vartheta\biglb(\omega-\Omega(q/n)\bigrb)}{\Gamma(q^2/2\pi^2n^2)}
      \,\frac{n}{D}\! \left[
                     \frac{\omega-\Omega(q/n)}{D}
                  \right]^{q^2/(2\pi^2n^2)-1}.
\end{equation}
Here $\vartheta(\omega)$ is the unit step function.  Its presence in
Eq.~(\ref{eq:structure_result}) expresses the obvious fact that the
minimum energy of a spin excitations with momentum $q$ is
$\varepsilon(q/n)$, Eq.~(\ref{eq:magnon_spectrum}).

The structure factor (\ref{eq:structure_definition}) is essentially a
Fourier transform of the correlation function $G_\perp(x,t)$ discussed
recently by Zvonarev \emph{et al.}\ \cite{zvonarev}.  Their treatment is
limited to the regime $q\ll n$; in which case our results are consistent
with Eqs.~(13) and (14) of Ref.~\onlinecite{zvonarev}.  On the other hand,
our calculations show interesting behavior at larger $q$, especially the
additional features at $\omega\ll J/\hbar$ and $q\approx\pm 2\pi n,
\pm4\pi n, \ldots$.

We now apply our technique to the calculation of the single-particle
spectral functions of the system
\begin{eqnarray}
  A^+_s(q,\omega)\!\!&=&\!\!\int\frac{dx\,dt}{2\pi}\, e^{-iqx+i\omega t}
                \langle\psi_s(x,t)\psi_s^\dagger(0,0)\rangle,
\label{eq:spectral_function_definition}
\\
  A^-_s(q,\omega)\!\!&=&\!\!\int\frac{dx\,dt}{2\pi}\, e^{-iqx+i\omega t}
                \langle\psi_s^\dagger(0,0)\psi_s(x,t)\rangle,
\end{eqnarray}
where $\psi_s(x)$ is the annihilation operator of bosons with spin
$s=\,\uparrow,\downarrow$.  

As discussed above, at strong repulsion ($\gamma\to+\infty$) the density
excitations of the system are identical to those of a gas of spinless
hard-core bosons, whose density $\Psi^\dagger(x)\Psi(x)$ equals the true
particle density $n(x)$.  (Here $\Psi$ is the annihilation operator of the
hard-core bosons.)  Then, assuming that the ground state is polarized in
the positive $z$-direction, one concludes that operator $\psi_\uparrow$
simply destroys a hard-core boson, i.e., $\psi_\uparrow(x)=\Psi(x)$.  In
the low-frequency regime $\omega\ll D$ the spectral functions
$A^\pm_\uparrow(q,\omega)$ can then be obtained in the framework of the
hydrodynamic approach based upon the Hamiltonian (\ref{eq:H_ph}) with
$K=1$.  In this method the annihilation operator $\Psi$ is expressed in
terms of the bosonic fields entering the Hamiltonian (\ref{eq:H_ph}) as
\begin{equation}
  \label{eq:bosonization}
  \Psi(x) = \sqrt{n}\, e^{-i\theta(x)} + \sqrt{n}\, 
            e^{-i\theta(x)}\sum_{j=1}^\infty
  [e^{i2\pi jl(x)} + e^{-i2\pi jl(x)}].
\end{equation}
Here one should use the hydrodynamic form (\ref{eq:l_bosonized}) of the
particle number operator $l(x)$.

Compared to the first term in the right-hand side of
Eq.~(\ref{eq:bosonization}), the remaining ones are formally irrelevant,
i.e., their contribution to the observable quantities is expected to show
additional power-law suppression at low energies.  The reason for writing
the full expression (\ref{eq:bosonization}) is that this form accounts for
the discreteness of particles by enforcing the condition of $l(x)$ being
integer \cite{haldane, giamarchi}.  As a result, at $\omega\ll D\sim n
u_\rho$ the spectral function $A_\uparrow(q,\omega)=
A^+_\uparrow(q,\omega) + A^-_\uparrow(q,\omega)$ shows not only the
expected feature near $q=0$, but also weaker features at $q=\pm 2\pi n,
\pm 4\pi n, \ldots$,
\begin{eqnarray}
A_\uparrow(q,\omega)&=&
\sum^\infty_{j=-\infty}\!\frac{\rho_\infty A_{|j|}}{\pi nu}
\frac{\Theta(\omega^2-u^2(q-2\pi jn)^2)}
{\Gamma((j-\frac12)^2)\Gamma((j+\frac12)^2)}
\nonumber\\
&&\times
\left(\frac{|\omega-u(q-2\pi jn)|}{2\pi nu}\right)^{\!(j-\frac12)^2-1}
\nonumber\\
&&\times
\left(\frac{|\omega+u(q-2\pi jn)|}{2\pi nu}\right)^{\!(j+\frac12)^2-1}.
\label{eq:spectral_function_up}
\end{eqnarray}
The hydrodynamic approach does not enable one to obtain the numerical
coefficients $\rho_\infty$ and $A_j$.  To find them, one can compare the
equal-time Green's function computed within this approach with the exact
results \cite{vaidya,jimbo,gangardt}.  This results in
$\rho_\infty=0.92418$, $A_0=1$, $A_1=1/16$, $A_2=9/2^{16}$,\ldots.

In this paper we are primarily interested in the spectral function
$A^+_\downarrow$, because unlike $A^\pm_\uparrow$, it is sensitive to the
non-trivial spin properties of the system. (The other spin-$\downarrow$
spectral function, $A^-_\downarrow$, obviously vanishes.)  To evaluate
$A^+_\downarrow(q,\omega)$, one needs to express the operator
$\psi_\downarrow(x)$ in terms of the density and spin variables entering
the Hamiltonians (\ref{eq:H_ph}) and (\ref{eq:Heisenberg}).  Following the
ideas of Refs.~\onlinecite{penc1} and \onlinecite{MFG} we identify
\begin{equation}
  \label{eq:annihilation_operator}
  \psi_\downarrow(x)=\Psi(x)Z_{l(x),\downarrow}.
\end{equation}
The presence of the hard-core boson operator $\Psi$ accounts for the
change in the total number of particles in the system, when a particle
with spin-$\downarrow$ is destroyed.  In addition, the number of sites in
the spin chain (\ref{eq:Heisenberg}) reduces by one.  This effect is
accounted for by the operator $Z_{l,\downarrow}$, which by definition
removes a site at position $l$ in the spin chain, provided that the spin
at that site is $\downarrow$.  (Otherwise, the result is zero.)

In the fully spin-polarized state of an infinite spin chain
(\ref{eq:Heisenberg}), the correlator $\langle
Z_{l\downarrow}Z_{l'\downarrow}^\dagger\rangle_\sigma$ coincides with the
spin-spin correlator (\ref{eq:spin-spin}).  Then the substitution of
Eq.~(\ref{eq:annihilation_operator}) into
(\ref{eq:spectral_function_definition}) gives
\begin{eqnarray}
  \label{eq:spectral_function_intermediate}
\hspace{-1em}
  A^+_\downarrow(q,\omega) &=& \int\frac{dx\,dt\,dQ}{(2\pi)^2}\,
                      e^{-iqx+i[\omega-\Omega(Q)] t}
\nonumber\\
                 &&\times\left\langle 
                       \Psi(x,t)e^{iQ[l(x,t)-l(0,0)]}\Psi^\dagger(0,0)
                         \right\rangle_{ph}.
\end{eqnarray}
To evaluate the expectation value in the ground state of the Hamiltonian
(\ref{eq:H_ph}), we use the hydrodynamic theory expression
(\ref{eq:bosonization}) for the hard-core boson operator.
Upon substitution of
Eq.~(\ref{eq:bosonization}) into
(\ref{eq:spectral_function_intermediate}), the effect of the $j\neq0$
terms amounts to the extension of the range of $Q$-integration from
($-\pi$, $\pi$) to ($-\infty$, $+\infty$).  Then the correlator in the
second line of Eq.~(\ref{eq:spectral_function_intermediate}) is computed
with the help of the relation
\begin{eqnarray}
  \label{eq:bosonization_greens_function}
  \left\langle
    e^{-i\theta(x,t)+iQl(x,t)}e^{i\theta(0,0)-iQl(0,0)}
  \right\rangle_{ph}  =
\nonumber\\
         \frac{e^{iQnx}}{
         [iD(t-x/u_\rho) + 1]^{\lambda^+_Q}
         [iD(t+x/u_\rho) +1]^{\lambda^-_Q}}
\end{eqnarray}
with $\lambda^\pm_Q=(Q/\pi\pm1)^2/4$, obtained using the standard
techniques \cite{giamarchi}.

Similarly to our derivation of the dynamic spin structure factor
(\ref{eq:structure_result}), at low frequencies one can neglect the
$x$-dependence in the denominator of
Eq.~(\ref{eq:bosonization_greens_function}) and find
\begin{eqnarray}
  \label{eq:spectral_function_result}
\hspace{-2em}
  A^+_\downarrow(q,\omega) &=&
      \frac{\vartheta\biglb(\omega-\Omega(q/n)\bigrb)}
           {\Gamma(q^2/2\pi^2n^2+1/2)}
\nonumber\\
      &&\times\frac{1}{D}\! \left[
                     \frac{\omega-\Omega(q/n)}{D}
                  \right]^{q^2/(2\pi^2n^2)-1/2}.
\end{eqnarray}
This expression is the main result of our paper.  It is worth noting, that
similarly to the case of $A_\uparrow(q,\omega)$,
Eq.~(\ref{eq:spectral_function_up}), the hydrodynamic approach does not
enable one to accurately determine the prefactor in
Eq.~(\ref{eq:spectral_function_result}), whose calculation at this time
remains an open problem.

The spectral function $A^+_\downarrow(q,\omega)$ is defined as the Fourier
transform (\ref{eq:spectral_function_definition}) of the spin-$\downarrow$
boson Green's function.  The latter was discussed recently by Akhanjee and
Tserkovnyak \cite{akhanjee}.  Their theory focused on the $Jt\to\infty$
limit, and accounted only for the long-wavelength magnons, $q\ll n$.
Calculating the inverse Fourier transform of
Eq.~(\ref{eq:spectral_function_result}) under these assumptions, we get
\begin{equation}
  \label{eq:greens_function}
  \langle\psi_\downarrow(x,t)\psi_\downarrow^\dagger(0,0)\rangle
  =\frac{n}{\sqrt{2\pi DJ/\hbar}}\, \frac{1}{it+0}
   \exp\left(\frac{i\hbar n^2x^2}{2Jt}\right).
\end{equation}
Comparison with the considerably more complicated Green's function given
by Eq.~(7) of Ref.~\onlinecite{akhanjee} shows the same oscillating
exponential factor (up to a missing $\pi^2$ in their exponent).  Further,
assuming $|x|\ll u_\rho t$ in the result of Ref.~\onlinecite{akhanjee}, we
find that their prefactor is consistent with our
Eq.~(\ref{eq:greens_function}).

It is interesting to compare the spectral function
(\ref{eq:spectral_function_result}) with that of strongly interacting
fermions \cite{MFG}.  The latter calculation, performed in the limit
$J\ll\hbar\omega$, shows the same Gaussian peak as a function of $q$ at
small $\omega$ as the expression (\ref{eq:spectral_function_result}) at
$\Omega\propto J\to0$.  In both cases the peak gives the leading
contribution to the density of states, obtained as $q$-integral of
$A^+_\downarrow(q, \omega)$, resulting in $\nu(\omega)\propto
1/\sqrt{\omega\ln (D/\omega)}$, cf.\ \cite{cheianov,fiete}.  In addition
to the peak at $q=0$, the spectral function of the fermion system shows
weaker features at the Fermi surface, $q=\pm k_F$, as well as the
shadow-band features at $\pm 3k_F$, $\pm5k_F$, etc., with the Fermi
momentum $k_F=\pi n/2$.  At $J/\omega\to0$ the boson spectral function
(\ref{eq:spectral_function_result}) does not show any additional features.
However, at $\omega \lesssim J/\hbar$ we find a sequence of additional
features at $q=\pm 2\pi n$, $\pm 4\pi n$, etc.

To summarize, we have developed a new approach to study the low-energy
properties of a gas of one-dimensional (iso)spin-$\frac12$ bosons with
strong short-range repulsion.  Our method is based on the separation of
density and spin variables in the form (\ref{eq:H_ph}) and
(\ref{eq:Heisenberg}) and the expression (\ref{eq:annihilation_operator})
for the boson annihilation operator.  We applied this technique to the
calculation of the dynamic spin structure factor
(\ref{eq:structure_result}) and the spectral function
(\ref{eq:spectral_function_result}).  At small $\omega$ they both show
Gaussian peaks as a function of $q$ centered at $q=0$, as well as
sequences of additional features at $q=\pm 2\pi n, \pm 4\pi n,\ldots$.

\begin{acknowledgments}
  The authors are grateful to T. Giamarchi, L. I. Glazman, G. V.
  Shlyapnikov, and M. B. Zvonarev for stimulating discussions.  K.A.M.  is
  grateful to RIKEN for hospitality.  This work was supported by the U.S.
  DOE, Office of Science, under Contract No.~\mbox{DE-AC02-06CH11357}, and
  by Grant-in-Aid for Scientific Research (Grant No.~16GS0219) from MEXT
  of Japan.
\end{acknowledgments}

\end{document}